# Beating the Heat! Automated Characterization of Piezoelectric Tubes for Starbugs

Rafal Piersiak*[a], Michael Goodwin[b], James Gilbert[b], Rolf Muller[b]

[a]Stony Brook University/Stony Brook, NY, USA; [b]Australian Astronomical Observatory, PO Box 915, North Ryde NSW 1670, Australia

## ABSTRACT

The Australian Astronomical Observatory has extensively prototyped a new robotic positioner to allow simultaneous positioning of optical fibers at the focal plane called 'Starbugs'. The Starbug devices each consist of two concentric piezoelectric tubes that 'walk' the optical fiber over the focal plane to accuracy of several microns. Ongoing research has led to the development of several Starbug prototypes, but lack of performance data has hampered further progress in the design of the Starbug positioners and the support equipment required to power and control them. Furthermore, Starbugs have been selected for the TAIPAN instrument, a prototype for MANIFEST on the GMT. A need now arises to measure and characterize 100's of piezoelectric tubes before full scale production of Starbugs for TAIPAN. The manual measurements of these piezoelectric tubes are a time consuming process taking several hours. Therefore, a versatile automated system is needed to measure and characterize these tubes in the laboratory before production of Starbugs. We have solved this problem with the design of an automated LabVIEW application that significantly reduces test times to several minutes. We present the various design aspects of the automation system and provide analyses of example piezoelectric tubes for Starbugs.

**Keywords:** Starbugs, fiber-positioner, automated, performance, characterization, piezoelectric, tube, d31

## 1. INTRODUCTION

Starbugs consist of two piezoelectric tubes (Figure 1c), one placed inside the other, that use a unique lift and step motion to travel across a glass field plate. They can move in four compass directions and have the ability to rotate in place. The applied voltages needed to produce any significant motion is on the order of 100-400V at frequencies of 50-300Hz. This requires the use of custom voltage amplifiers to achieve such high voltages. Starbugs are miniature piezoelectric robots that can quickly and accurately position many optical 'payloads' (e.g. fibers) simultaneously. They were first conceptualized in 2004 [8], and later various prototypes and mounts are described in 2006 [5] along with the Starbugs locomotion technique in 2010 [3].

The characterization data utilized in this paper was attained in 2012, while [1] was being written. Great progress on Starbugs eventually lead to serious discussion of tube design for MANIFEST (the Many Instrument Fiber System), which requires various tube models varying in diameter and weight [2]. The results in this paper present some key parameters that will decide what tubes are coupled together to perform a specific task. Starbugs is currently scheduled for deployment with the TAIPAN (Transforming Astronomical Imaging-surveys through Polychromatic Analysis of Nebulae) spectroscope on the UK Schmidt Telescope in 2015, where 150 Starbugs will be independently carrying a fiber bundle [7]. The automated characterization system developed in this paper will be repurposed for quality control and functionality tests on the Starbugs destined for TAIPAN.

Current positioning technology, such as 2dF on the Anglo-Australian Telescope (AAT) is limited in how fast it can configure a plate for observation. It is slow, requires two field plates (one is used for observation, while the second is reconfigured for the next observation), has limited light capture, and can't reposition fibers without obstructing others. It is common for 2dF to take over one hour to sequentially reposition 400 fibers. By contrast, Starbugs is limited by the furthest traveling Starbug, which may require a few minutes to reach its final destination.

Although, 2dF is a sufficient technology for the 4m AAT, it would be highly impractical for Extremely Large Telescopes (ELT's), such as the proposed 24.5m Giant Magellan Telescope (GMT). A scaled up version of 2dF for the GMT would contain well over 1000 fibers, which would all be positioned serially. The linearity associated with this instrument severely limits how many observations can be taken on any particular night. The parallel nature of Starbugs removes this linear restriction and makes it high scalable. Starbugs will be part of MANIFEST, which will be utilized on the GMT [9].

*rafal.piersiak@alumni.stonybrook.edu; www.aao.gov.au

Starbugs eliminates many of these problems by providing each miniature robot with its own propulsion system. They can move simultaneously to their respective targets, carrying a wide variety fiber bundles with large light collecting areas. Other merits include the ability to track targets in real-time without light obstruction.

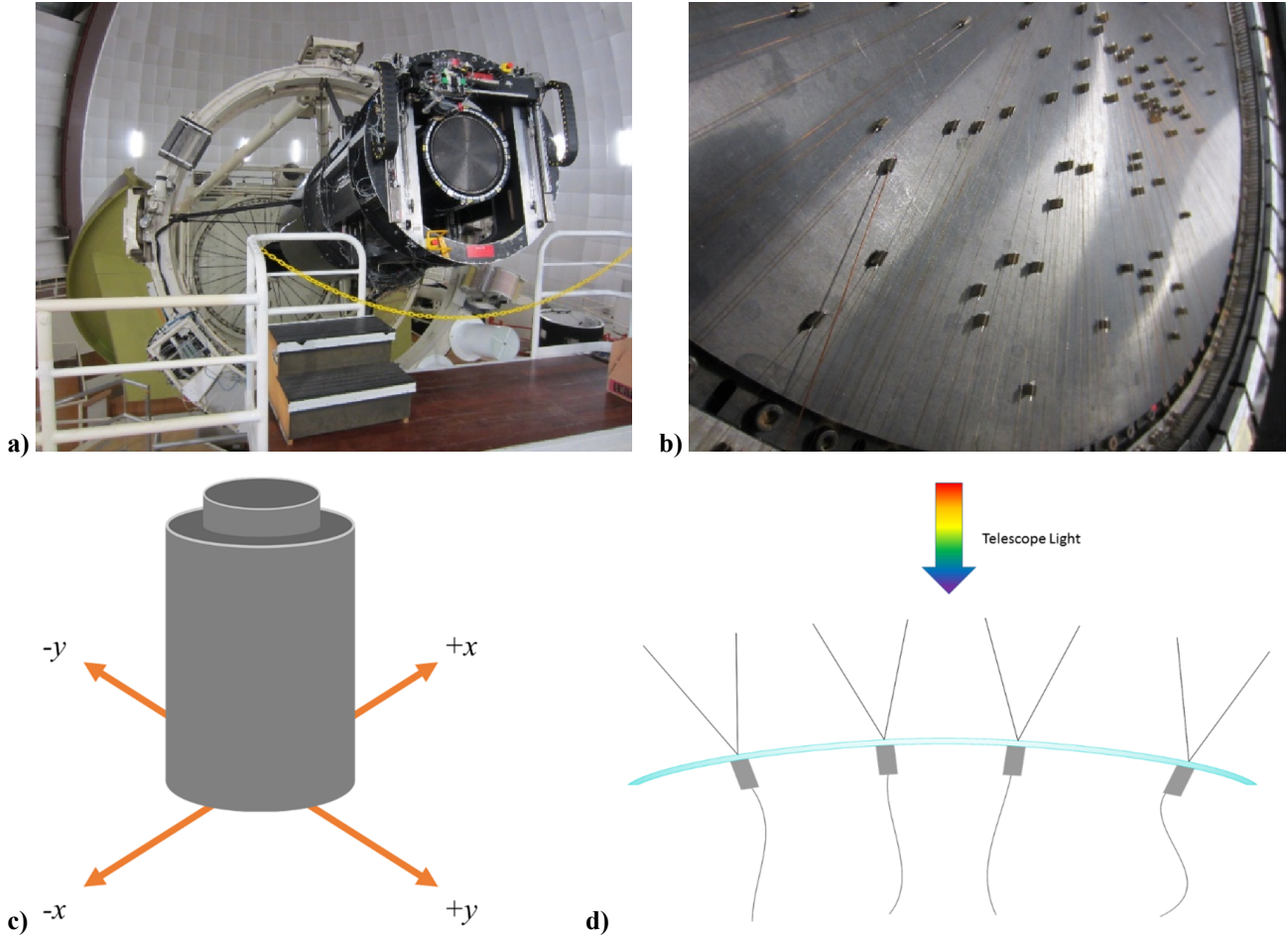

Figure 1. a) 2dF Robotic Positioner on the AAT sequentially deploying 400 fibers on a 1.3m metal field plate; b) Magnetic fiber buttons positioned on the 2dF field plate; c) The two concentric tubes of a Starbug use a unique motion to move in the ±x and ±y direction. The central aperture can accommodate a variety of fiber hexabundles used for collecting light; d) Starbugs hang from a glass field plate by vacuum suction and collect light, which is fed to a spectrograph using fiber optic bundles.

## 2. NEED FOR PERFORMANCE DATA

The power requirements of a single piezoelectric tube can be massive, on the order of 50W at full power. Factor in that there are two tubes and you can see why Starbugs need to be carefully designed to limit excessive power consumption. There is also a trade-off between power consumption and step size. Achieving a larger step size requires more power. Currently there is a lack of data that correlates these two factors in a way that can be used in the design process to achieve an optimal condition of reasonable power dissipation and step size.

Gathering a sample of data manually using a function generator and oscilloscope for one tube exceeds a reasonable amount of time for such a task, on the order of several hours. Although the logged data was insightful, the time and effort associated with the task was not efficient. The logical next step was to develop a test application to automatically test a tube and store the data, allowing the engineer to make progress on other tasks in the interim.

The development of this stand-alone automated test application exemplifies a simplistic, natural workflow, critical to its ability to process a large amount of piezoelectric tubes swiftly and economically. The goal was to log power, temperature, and the axial and lateral deflection of the piezoelectric tube over a variety of voltage and frequency combinations. The data will be processed and displayed graphically using a MATLAB script to generate plots and 3-D graphs, showing how various parameters changed over time and with respect to voltage and frequency. MATLAB saves these plots and graphs into easy to read reports, which can then be used to parameterize various piezoelectric tubes. The designer can use these parameters to select the best tube for a particular environment.

## 3. LABORATORY TESTING

The automated data acquisition setup is comprised of LabVIEW software and hardware and .csv (comma-separated values) files, which will be used for characterizing piezoelectric tubes.

### 3.1 Experimental Setup

The automated test application required the use of two specialized instruments along with a laptop computer with LabVIEW installed. The laptop is loaded with a custom LabVIEW application, which carries out the actual test and data logging utilizing National Instruments (NI) modules as an interface between the laptop and piezoelectric tube. The setup uses four modules: ±10V output voltage module (9263), thermocouple module (9214), high voltage input module (9225), and a high current input module (9227). These modules work together to excite the piezoelectric tube and gather data. The Starbugs high voltage amplifier was a custom design built for Starbugs characterization. It takes in a ±10V input signal and amplifies it 20x, with a peak output voltage of ±200V. The piezoelectric tubes have inner and outer electrodes with wires soldered onto them to provide a path for current. A thermocouple is placed against the inside of the tube using a brass metal strip which is wound up and acts as a spring, pressing the thermocouple against the inner wall of the piezoelectric tube.

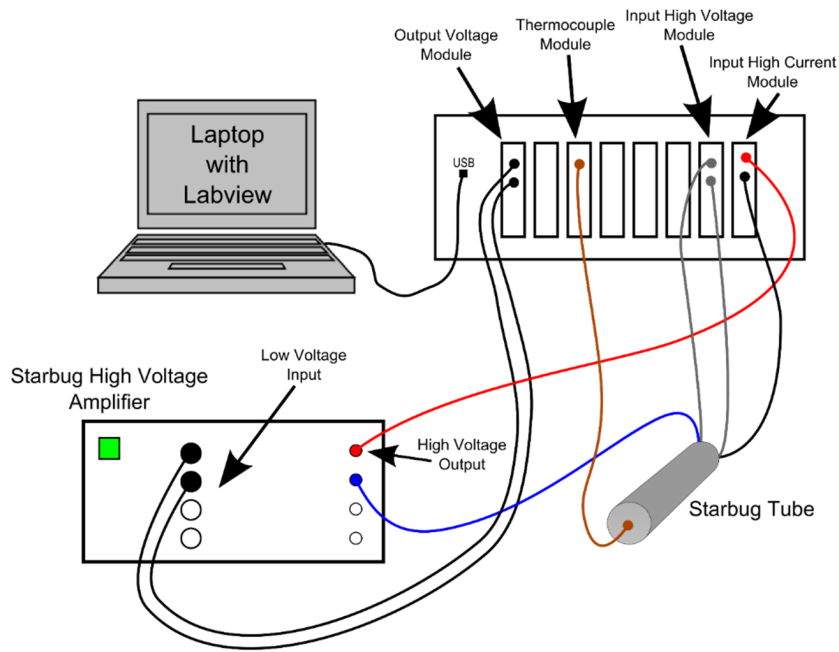

Figure 2: This is a schematic representation of the bench top used for running the automated test application for characterizing piezoelectric tubes. There are four main components: A laptop with LabVIEW installed, a compact Data Acquisition Unit (cDAQ) from National Instruments, a custom built high voltage amplifier, and a variety of piezoelectric tubes.

The hardware setup found in (Figure 3) shows the NI cDAQ and four cDAQ modules connected to a piezoelectric tube. A type T thermocouple was used to measure the temperature of the piezoelectric tube. The piezoelectric tube is lifted in the air to simulate similar conditions it would experience while hanging underneath a glass field plate.

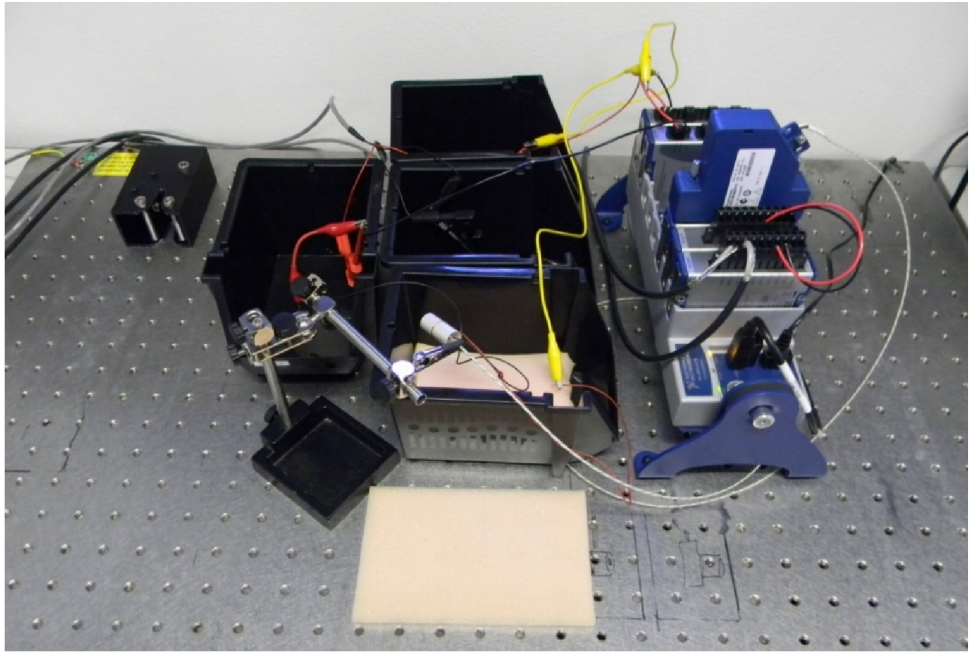

Figure 3: Experimental setup used to measure the power and temperature of a piezoelectric tube during an automated test.

### 3.2 Deflection Setup

Starbugs move using two methods of mechanical movement. They can expand lengthwise, known as axial deflection, seen in (Figure 4a) or they can tilt sideways, called lateral deflection (scan range), seen in (Figure 4b). The outer tube on a Starbug utilizes axial deflection to lift the inner tube off the surface, while the inner tube utilizes lateral deflection to move the entire Starbug. These two deflections work together to create a "lift and step" motion [3]. Larger axial and lateral deflections allow the tubes to move at a quicker pace.

The deflection test setup (Figure 5) required a bit of ingenuity and some blocks borrowed from the photonics industry. A capacitive sensor is mounted on a platform and held in place by a clamp. A piezoelectric tube with a metal plate and a plastic isolation shield (to prevent conduction) attached to one end is mounted on another platform and is lightly clamped in place.

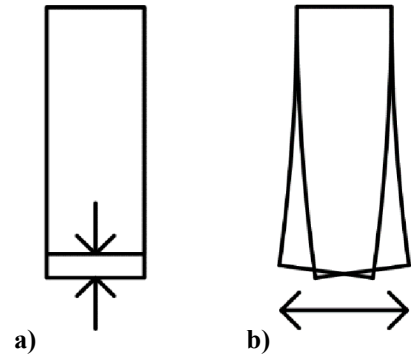

Figure 4: a) Axial deflection occurs when the outer wall is applied with a positive voltage, while the inner wall is brought to ground or is applied with a corresponding negative voltage; b) Lateral deflection occurs when one of four outer electrodes is applied with a positive voltage, while the opposite facing outer electrode is brought to ground or is applied with a corresponding negative voltage.

Capacitance is inversely proportional to the distance between two plates. The sensor uses this property to detect a change in capacitance between itself and the metal plate by capacitively measuring the distance. The sensor connects to an instrument, which converts the change in capacitance to a waveform. Measuring the peak-to-peak voltage of the waveform and dividing it by the instrument's intrinsic sensitivity gives you the displacement. The displacement can then be used to calculate the "$d_{31}$" constant, which parameterizes the relative change in length of a piezoelectric tube.

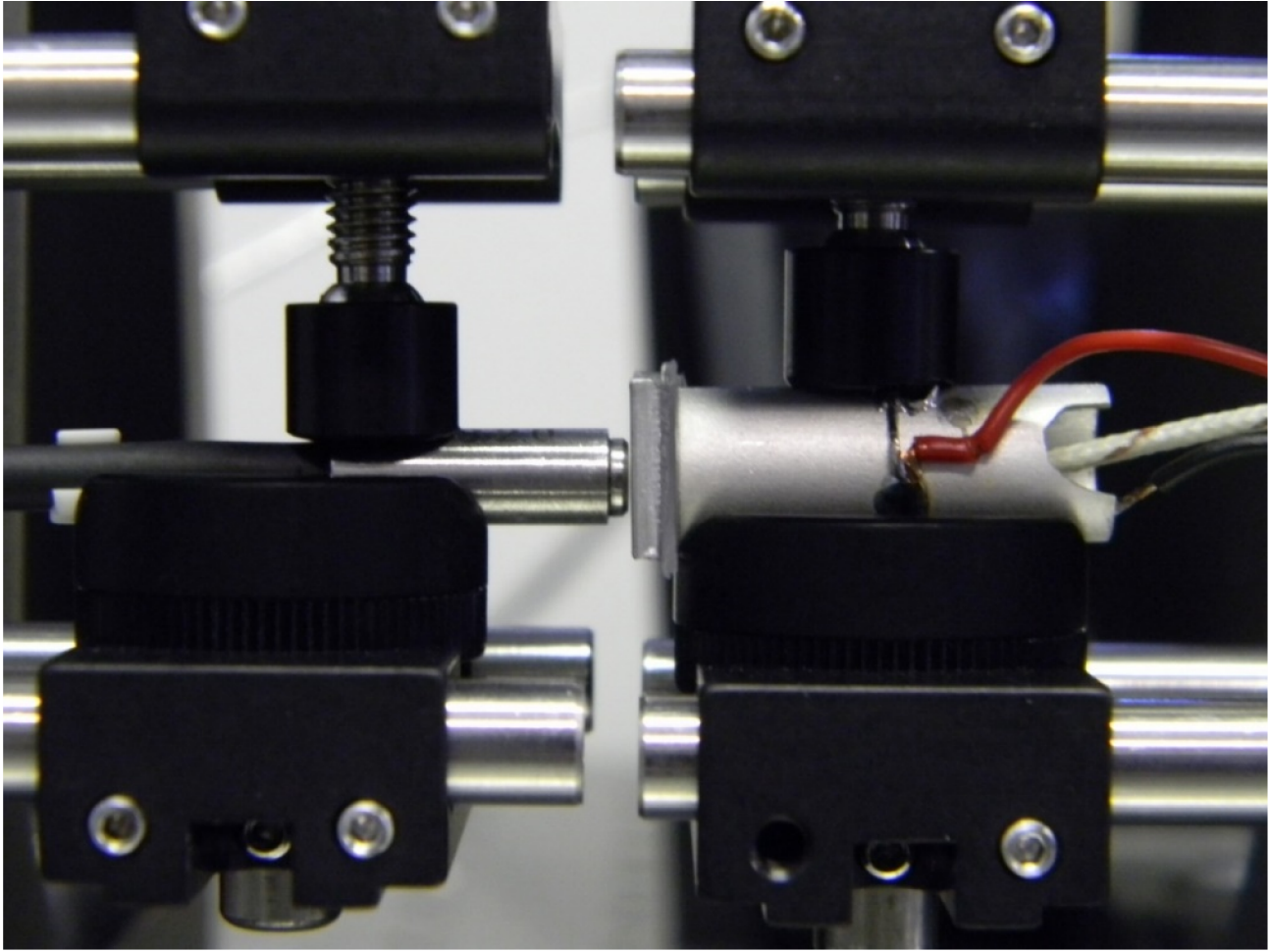

Figure 5: Axial deflection test setup using a capacitive sensor and a piezoelectric tube with a metal plate and plastic isolation shield attached to one end. A thermocouple is also placed inside the piezoelectric tube. A similar setup can be recreated for a lateral deflection test. A metal plate and plastic shield would have to be mounted to the outer wall on one end of the piezoelectric tube. The capacitive sensor would be placed, facing the metal plate mounted on the outer wall on the end of the piezoelectric tube.

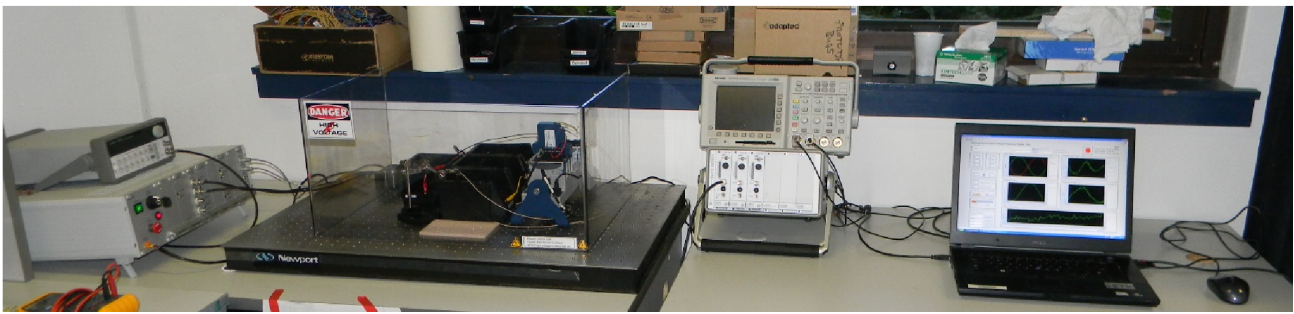

Figure 6: Starbugs automated characterization setup (left to right: function generator, custom high voltage amplifier, Starbugs power/temperature setup, oscilloscope, capacitive displacement sensor instrument, LabVIEW front panel).

# 4. DATA AQUISITION

The test application was programmed using NI's LabVIEW software. LabVIEW uses a graphical programming language to create a virtual test instrument. It allows the designer to display graphs, create switches, and provide numeric inputs on a front panel. The user interacts with this instrument using the front panel. An example of LabVIEW code can be found in (Figure 9). The Voltage Generation and Sweep code clock utilizes a finite-state machine (FSM) to control dataflow, while the Progress Bar/Timing and Data Logging code blocks follow a conventional LabVIEW dataflow path.

## 4.1 LabVIEW Dataflow Code Diagram

The Voltage Generation and Sweep block takes care of generating a sinusoidal signal with a variable amplitude and frequency. It implements a sweep algorithm which changes the voltage and/or frequency based on the initial and final parameters set by the user. The user can specify exactly how long each voltage/frequency combination should be output. This block also checks for errors, reads inputs, waits for the start command, and sends a "Test Complete" email and plays a sound when the test has been successfully completed.

The Progress Bar block initializes the progress bar and updates it continuously. It also displays the elapsed time and calculates the expected test duration.

The Data Logging block initializes and reads the input modules, performs calculations, and writes to a spread sheet. The data is stored in an organized .csv format. It also formats the file names and monitors the thermocouple for the peak temperature set by the user. If the thermocouple reads a value equal or greater than the set peak temperature, the output signal will be disabled, an error email will be sent, an error sound will play, and the program will terminate. This feature prevents the piezoelectric tube from reaching its Curie temperature and depoling.

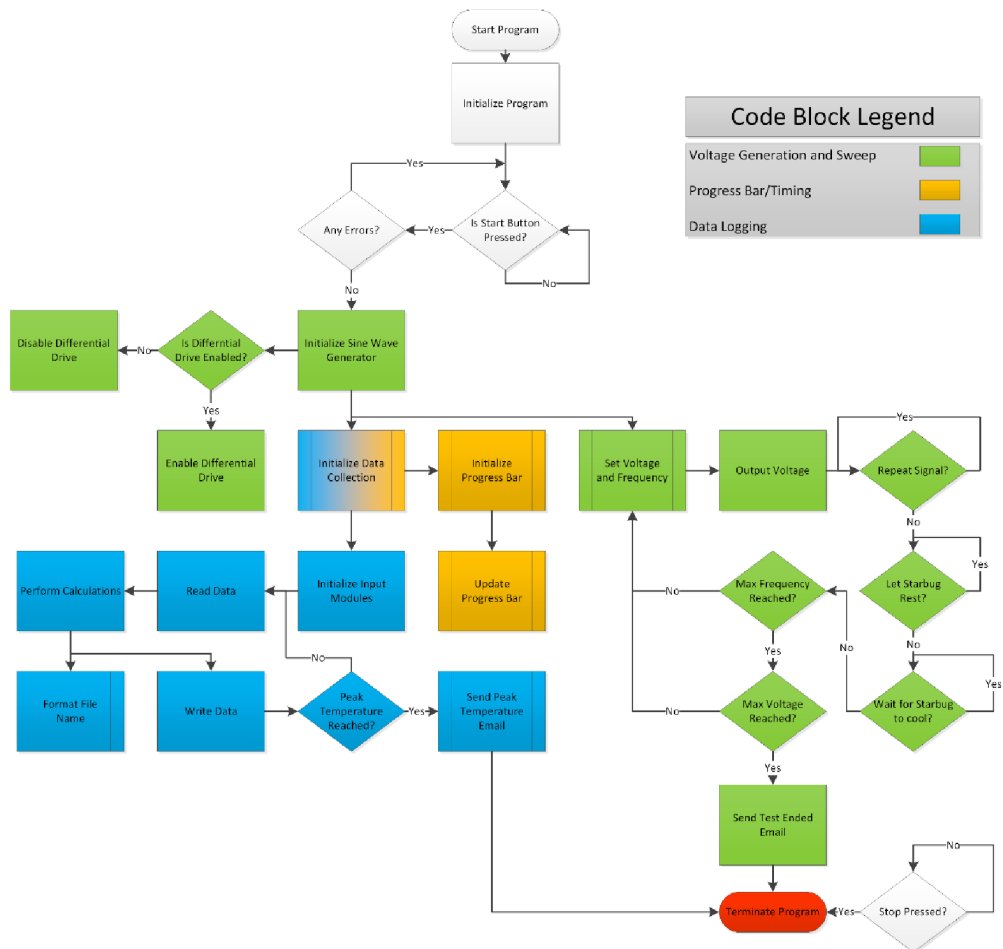

Figure 7: Summary of the dataflow and decisions the automated LabVIEW test application makes.

The simplest approach to create a "sweep" is to use a FSM implemented to simplify decision making. An FSM is comprised of "states", which perform various functions. An FSM begins in an initial state and waits for a certain condition or period of time before it transitions to the next state. The following (Figure 8) displays a state diagram, which is a pictorial representation of the actual implementation of the FSM used in the Voltage Generation and Sweep block.

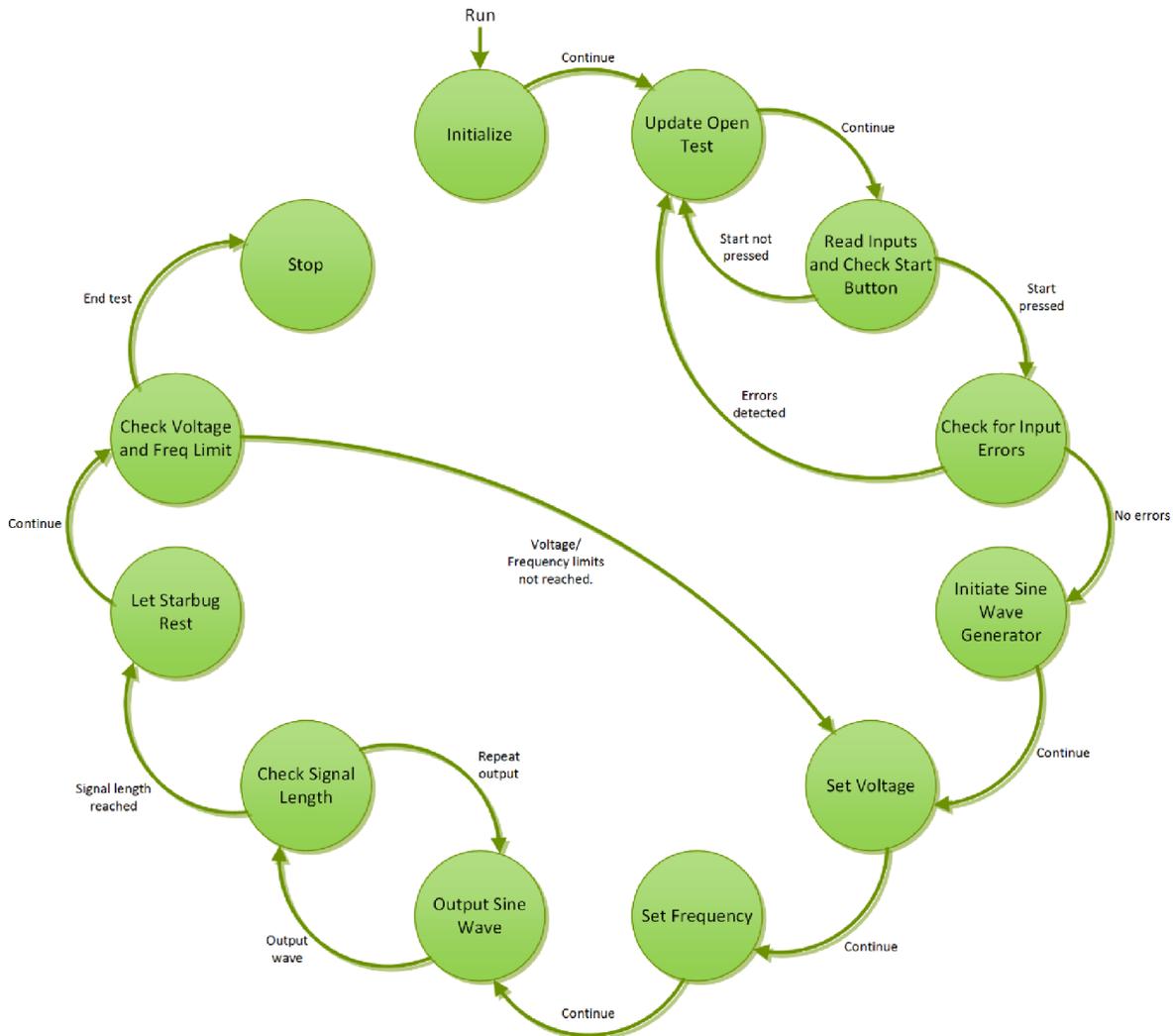

Figure 8: State Diagram of the FSM used in the Voltage Generation and Sweep block. The FSM starts at the "Initialize" state and continues until it has reached the "Stop" state.

### 4.2 LabVIEW Code

The program starts by initializing several functions and waits for the user to press the start button. Once pressed the program checks for errors and continues onwards if no faults are detected. It proceeds by initializing the output voltage module, which sets the stage for enabling the three main blocks of code. While the sinusoidal wave is being configured, the input modules are initialized. The sinusoidal wave is sent out and then the Data Logging code block begins to read the input modules, perform calculations, and write the date, while checking if the user-defined peak temperature was reached. After the write, the Voltage Generation and Sweep block checks if the signal should be repeated. If not, it checks to see if it should wait before outputting any other waveforms. If not, it checks whether it should wait until the piezoelectric tube has cooled to a set temperature. Once it completes this task, it verifies if it has run through each voltage and frequency combination. If it has, the program will send an email notifying the user the test has successfully completed and the application will terminate.

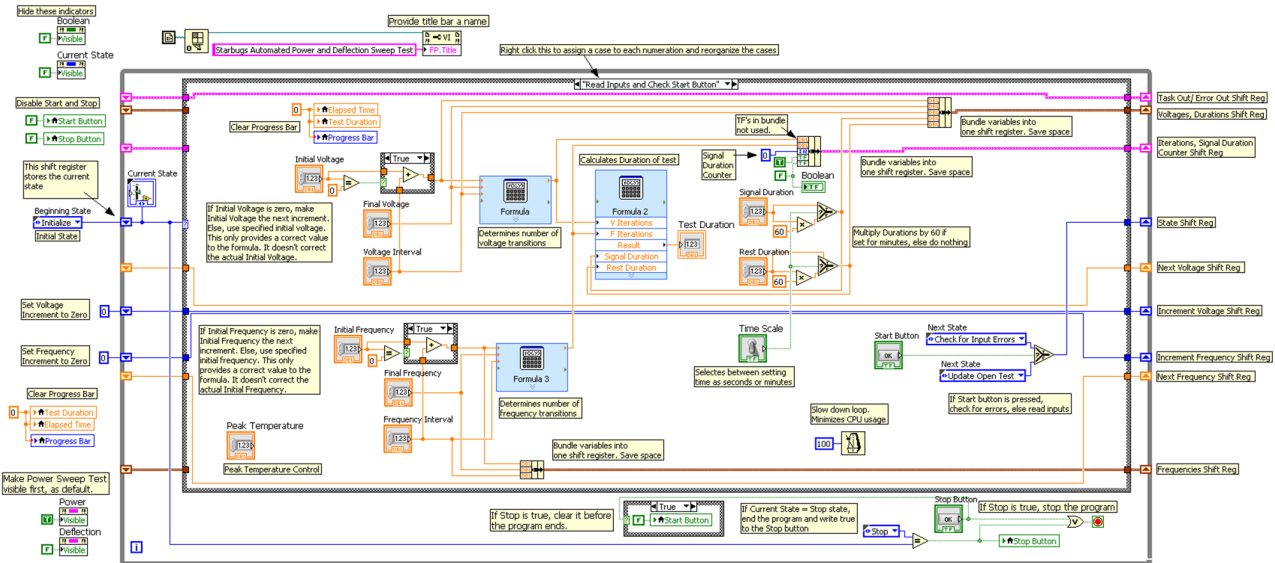

Figure 9: Example of one of the states for the Voltage Generation and Sweep code block

## 4.3 Automated Test Application Front Panel

This seemingly complicated code boils down to a clean and intuitive user interface called a "Front Panel." The user interacts with the test application by modifying the controls found on the front panel, such as switches, tick boxes, and numeric controls. Data is graphed in real-time so the user can actually see some of the data being recorded, while the test is running. The following two figures show the front panel generated by the LabVIEW code. The user enters various voltage, frequency, timing, and temperature parameters and presses start. An email will notify the user that the test has completed successfully, whether peak temperature has been reached, and various trouble errors.

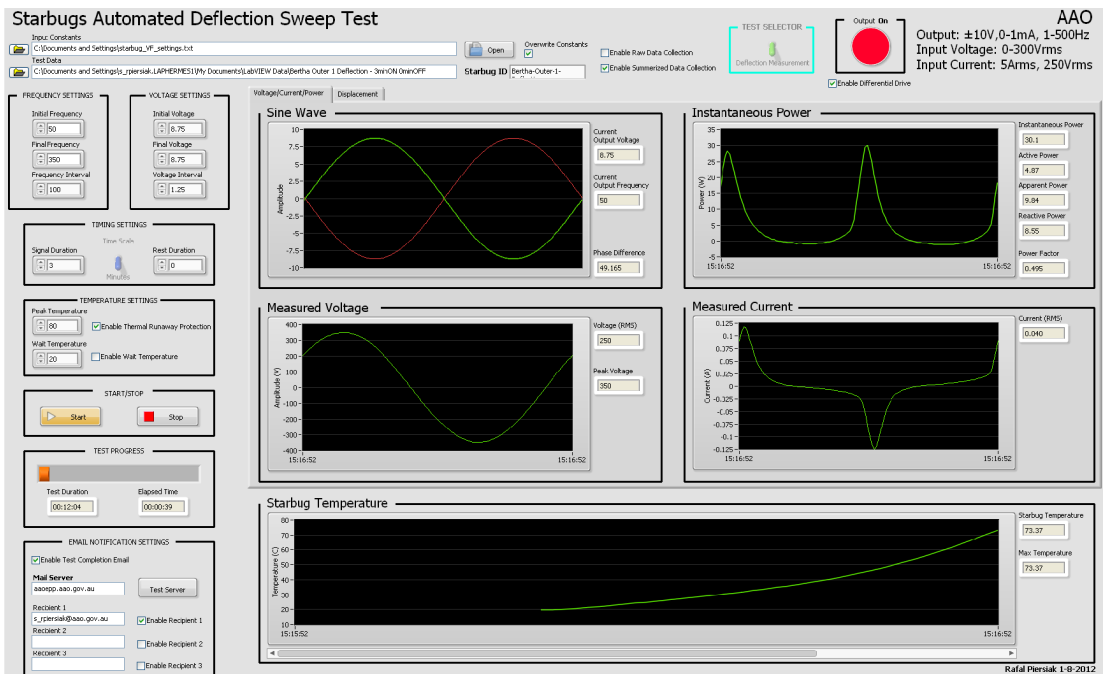

Figure 10: This screenshot shows a live test being performed on a piezoelectric tube. The four upper graphs, going clockwise from the top left display the output voltage, instantaneous power, measured current through the piezoelectric tube, and the measured voltage across the piezoelectric tube. The bottom graph displays the piezoelectric tube temperature.

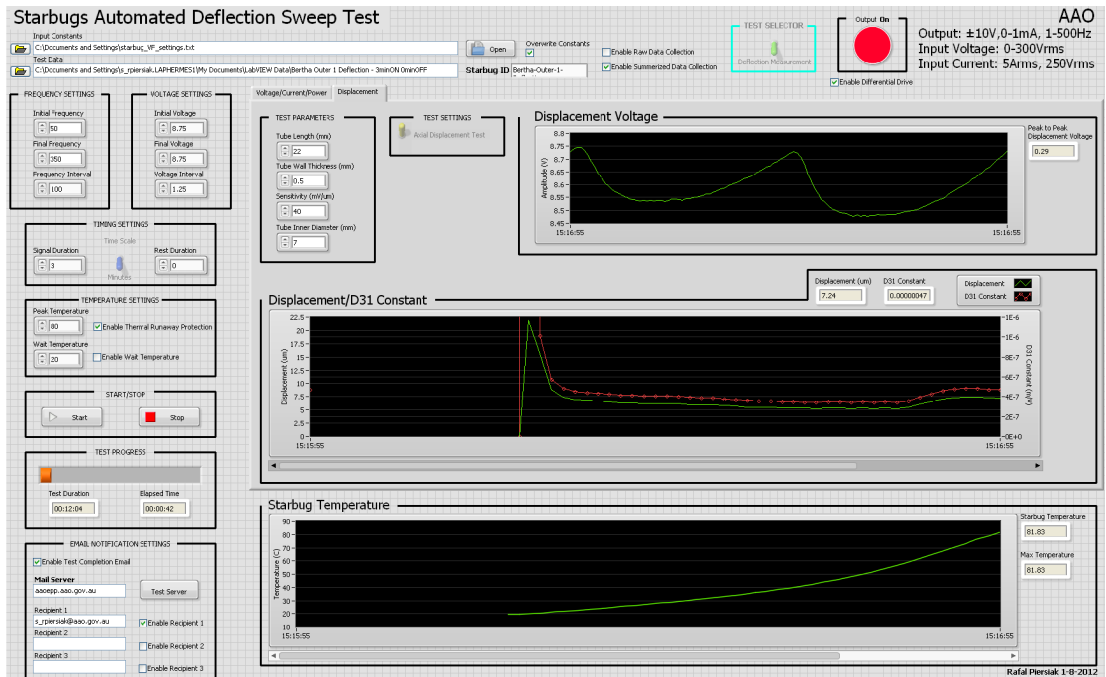

Figure 11: This is another screenshot of the same test performed in the previous figure. The top graph shows the voltage output by the capacitive sensor instrument. The middle graph shows both the displacement (green) of the piezoelectric tube and the corresponding $d_{31}$ constant (red) on separate scales.

### 4.4 Data Storage and Format

The LabVIEW test application writes data in the .csv format and organizes each parameter in a separate column. Values are stored for each second a waveform is applied to the piezoelectric tube. To limit the size of the files, sampling has been reduced to once per second. Faster sample times easily create complete datasets of ~500 MB, which can be difficult to process and does not offer added insight.

## 5. PERFORMANCE

The results in this section are for the prototype Starbugs piezoelectric tube, "Bertha Outer". Five key areas of performance are discussed: i) capacitance; ii) power; iii) temperature; iv); deflection and v) active cooling [4] [6].

### 5.1 MATLAB Plots

A script was developed to read the .csv files generated by LabVIEW and plot each parameter over time. This provides a quick summary of what is happening to a piezoelectric tube for a sinusoidal wave at a particular voltage and frequency. It graphs the following parameters over time: Starbug temperature, RMS measured voltage, RMS measured current, active power, apparent power, reactive power, power factor, phase angle, impedance, resistance, reactance, capacitance, loss tangent, peak to peak displacement voltage, displacement, axial $d_{31}$/lateral displacement. These plots don't show how the tube reacts to different sinusoidal waveforms, so another script was developed using Delaunay triangulation along with interpolation to create 3-D plots as a function of voltage and frequency.

### 5.2 Thermal Properties

Piezoelectric material has a Curie point, which defines the temperature where demagnetization of the material occurs. Once the material becomes paramagnetic, only an electric force can cause its magnetic field to align in an ordered manner. At this point any magnetic properties of the piezoelectric material are compromised, reducing the material into a variable resistor. The piezoelectric material used for the Starbugs tubes has a Curie point of 80°C. Surpassing this temperature for more than 2-5 seconds destroys the tube, as lab tests have shown.

Piezoelectric tubes have a peculiar thermal property that was revealed when a thermal camera was used to ensure the thermocouple was placed near the hottest section of the piezoelectric tube. While the inner wall becomes hot (~70°C) near the edges, the outer wall stays at a very comfortable temperature ranging from 20-25°C. The piezoelectric material has the unexpected property of insulating the outer wall and keeping the temperature well below temperatures measured inside the tube. The Starbugs design doesn't expose the inner walls to the environment. And because the outer walls don't transfer heat from the inner walls to the surrounding environment, the prospect of fan cooling is restricted.

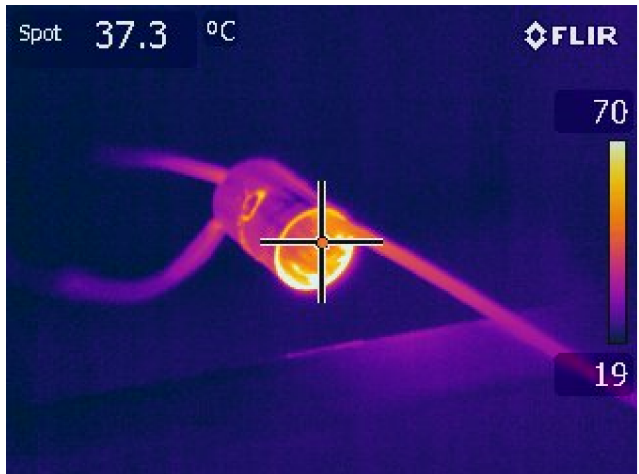

Figure 12: Betsy Outer internal thermal probe

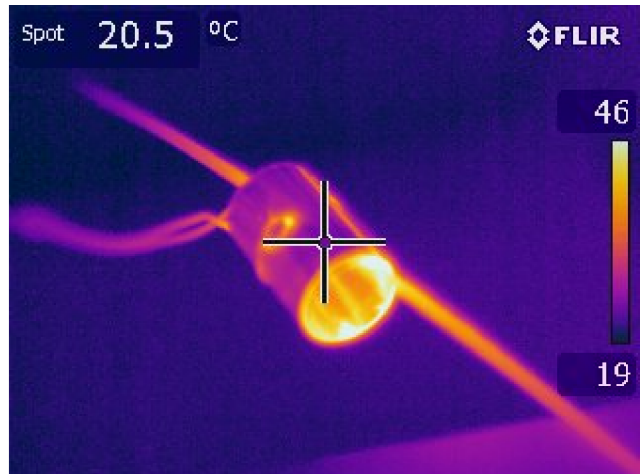

Figure 13: Betsy Outer external thermal probe

### 5.3 Automated Characterization Specifications

The main representative data analysis will be performed with the Bertha Outer tube. The dimensions are: 22mm long, inner diameter of 7mm, and a wall thickness of 0.5mm. It is assumed a Starbug will not travel for more than three minutes during initial configuration on MANIFEST [1]. Therefore an automated test will advance for 3 minutes for each voltage/frequency combination. There was no wait period for the tube to cool to a specified temperature between voltage/frequency combinations, but the option is available in the front panel. Generally, temperature builds up gradually with frequency at each voltage output, so increasing frequency won't disturb the final stable temperate (if it exists). Skipping the cooling period speeds up the test speed. The room temperature was approximately 19°C, which can be validated by reading the starting temperature at the initial voltage/frequency output combination, assuming the tube has had time to cool from a previous automated test.

Reports have been created for all tested tubes outlined in Table 1. Tubes with "Outer" as a location are used as the outer tube, which experiences axial deflection. "Inner" tubes experience lateral deflection. Each "Outer" and "Inner" complement each other and create a Starbug. The tube named "Mohawk" is being used for the upcoming DESpec spine positioner, which only requires one 'single' tube [10]. A number appended to the tube name and location denotes priority for testing purposes only.

### 5.4 Capacitance

Piezoelectric material is naturally capacitive and responds differently to variously frequencies. Maintaining low power operation requires a low capacitive load in order to minimize the dissipative loss tangent. The capacitance is low (~50nF) and quite stable in the region defined as the 'sweet spot' below, up to 350Hz at 50-200V and up to 250Hz at 250V.

Any attempts to operate outside these regions will result in a huge gain in capacitance, which consequently causes an undesirably large increase in power. The increasing capacitance of this normally capacitive device eventually causes a chain reaction of events that if not monitored, will demagnetize the piezoelectric material and render it unusable.

Table 1: The capacitance is extracted for the data at ($T_{end}$ – 1) seconds, normally 179 seconds to avoid false data. The voltage and frequency columns define the extent of the sweet spot, where the tabulated voltage/frequency combination for each tube represents the max point of controlled thermal dissipation.

| Tube Name | Capacitance (nF) | Voltage (V) | Frequency (Hz) |
|---|---|---|---|
| Bertha Outer 1 | 75 | 250 | 250 |
| Bertha Inner 6 | 54 | 200 | 250 |
| Betsy Outer 2 | 127 | 200 | 250 |
| Betsy Outer 4 | 84 | 250 | 250 |
| Belinda Outer 5 | 30 | 350 | 150 |
| Belinda Inner 7 | 12 | 350 | 250 |
| Mohawk Single 3 | 31 | 300 | 150 |

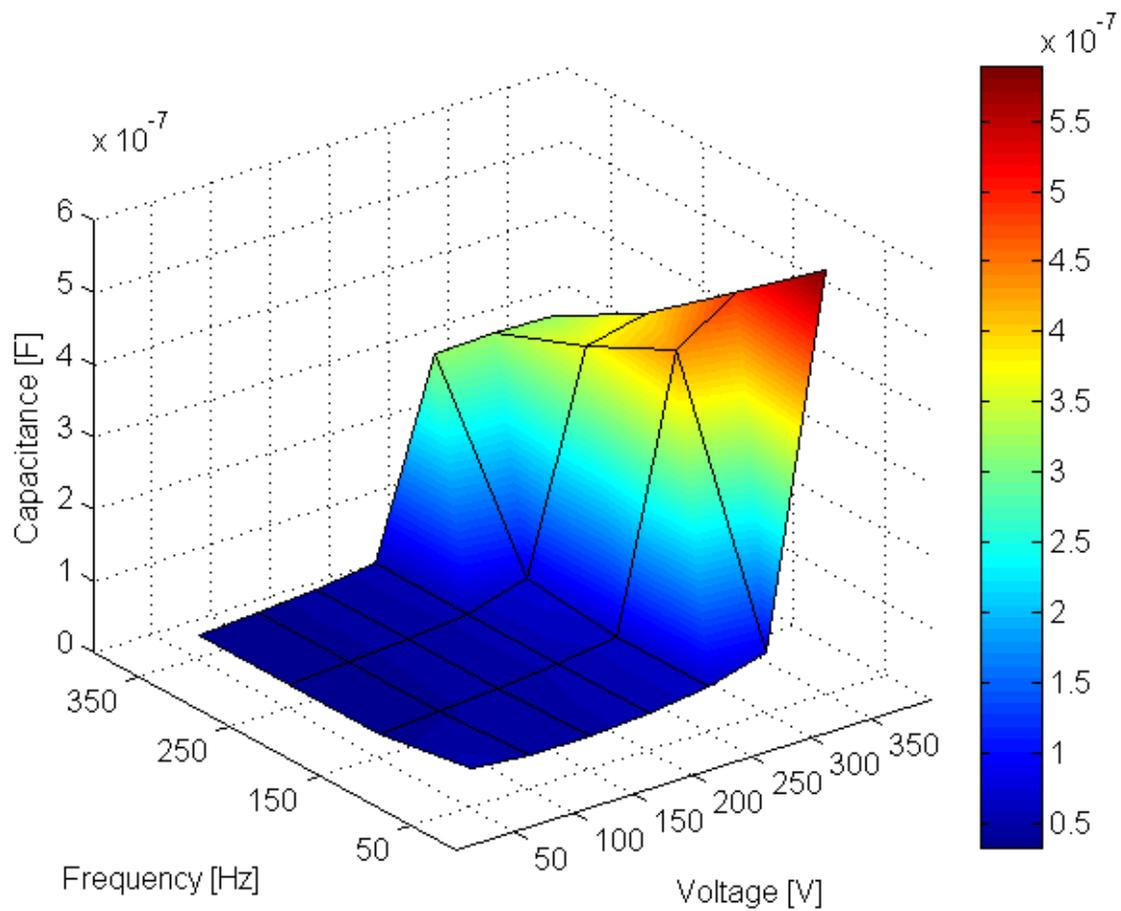

Figure 14: Capacitance surface graph created with data from an automated test with Bertha Outer. The sweet spot is in the region up to 250V and 250Hz, but can be pushed to 300V (and up to 250Hz) if power dissipation and temperature are not an issue. At very high voltages and frequencies, the capacitance slowly becomes smaller.

## 5.5 Power

As the capacitive plot has predicted, the active power defines the same sweet spot region of practical and safe operation. Outside this threshold, power dissipation jumps drastically from 1W to upwards of 17W. Running a piezoelectric tube under such high power conditions will not only lead to disruptive demagnetizing temperatures, but will bring significant stress on the power amplifiers used to power these piezoelectric tubes.

The capacitance begins to decay somewhat at very high voltages and frequencies, but the active power continues to rise, showing there is only moderate correlation between capacitance and increased active power. The demagnetization of the piezoelectric tube transforms the material to a more resistive (vs. capacitive) device as the power factor increases, causing the reactance to drop; lowering capacitance at the high end (350Hz, 350V), while resistive forces take precedence.

Comparing the slopes (voltage and frequency) of the active power reveals that a change in voltage causes a significant increase in active power versus frequency, which clearly follows the definition for dynamic power. Based on modulation of voltage/frequency, no assumption can be made on mechanical performance with purely electrical characterization.

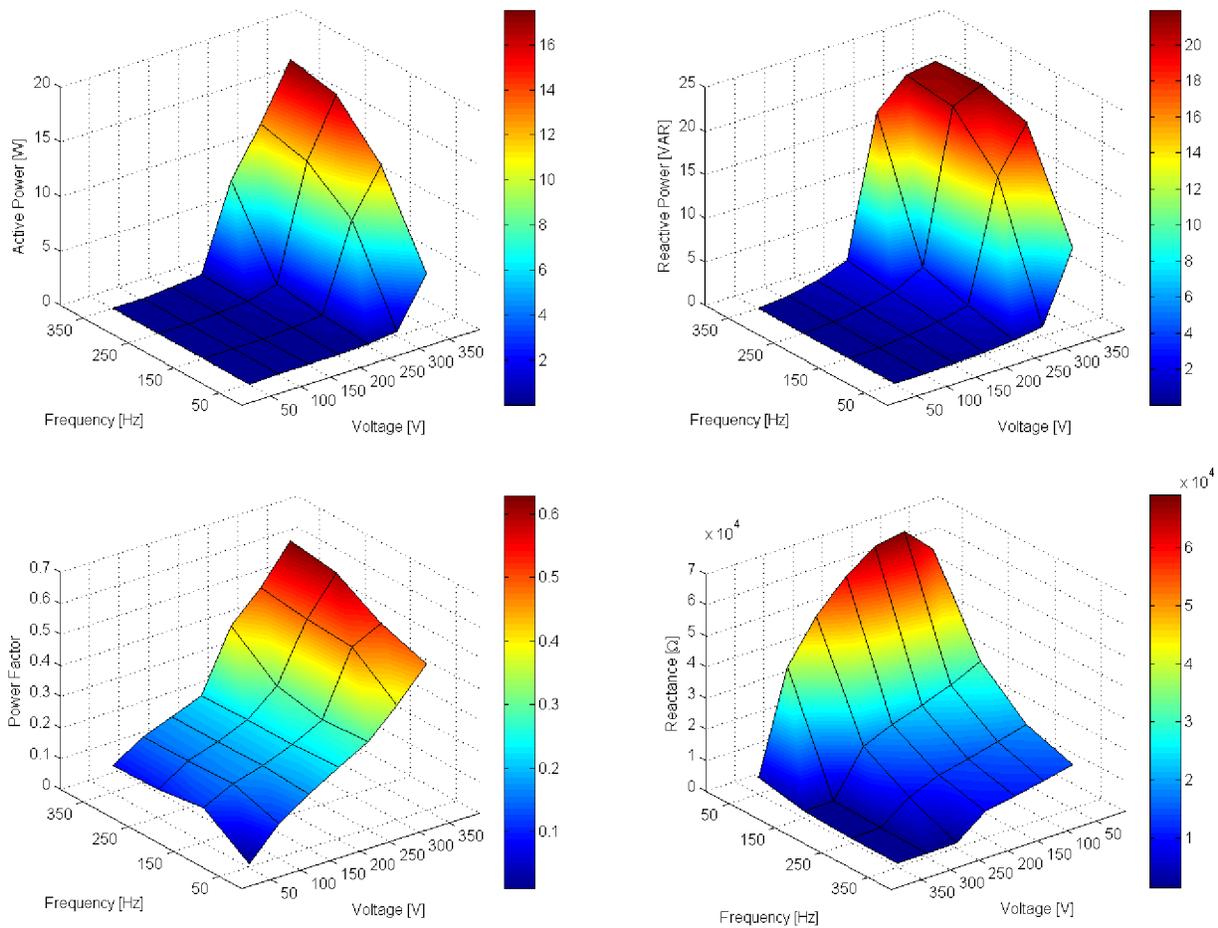

Figure 15: Power characteristics plotted with data from an automated test with the Bertha Outer piezoelectric tube. (From top left, continuing clockwise: active power, reactive power, power factor, and reactance (rotated 180°) vs. voltage/frequency respectively.

Table 2: The parameters are extracted for the data at ($T_{end}$ − 1) seconds, normally 179 seconds to avoid false data. Each parameter is based on the max point of controlled thermal dissipation.

| Tube Name | Active Power (W) | Reactive Power (VAR) | Power Factor | Reactance (Ω) |
|---|---|---|---|---|
| Bertha Outer 1 | 1.04 | 3.30 | 0.30 | 8500 |
| Bertha Inner 6 | 0.38 | 1.60 | 0.24 | 11900 |
| Betsy Outer 2 | 0.77 | 3.83 | 0.20 | 5000 |
| Betsy Outer 4 | 1.19 | 3.70 | 0.31 | 7600 |
| Belinda Outer 5 | 0.60 | 1.46 | 0.38 | 35200 |
| Belinda Inner 7 | 0.34 | 1.00 | 0.32 | 54200 |
| Mohawk Single 3 | 0.47 | 1.14 | 0.38 | 33800 |

### 5.6 Temperature

Piezoelectric material is extremely sensitive to temperature. Producing high voltage/frequency (e.g. 350V/350Hz) combinations common for Starbugs can lead to thermal runaway. Piezoelectric tubes can produce a temperature difference of 45°C in four seconds; surpassing the Curie point generates irreversible damage and can lead to some explosive fireworks as well as a major fire hazard. This study is critical to not only design an efficient Starbugs system, but also an innocuous one.

Any attempt to operate the piezoelectric tube outside the sweet spot found in the following figure will result in thermal runaway and damage the tube. The Bertha Outer tubes can be operated at 250V up to 250Hz at stable temperatures. Going beyond these initial conditions will increase power exponentially and can certainly lead to temperatures excessive of the Curie point. The same holds true for other Starbugs piezoelectric tubes, although the sweet spot could extend or retreat.

A more intuitive method of predicating thermal runaway is through the loss tangent, which takes the ratio of active power vs. reactive power. The loss tangent explains how effective a dielectric material is at dissipating heat. As the loss tangent increases, the dielectric material becomes less efficient at extruding heat from itself, causing substantial heat buildup that cannot be dispelled into the surrounding medium. Bertha piezoelectric tubes are stable with a loss tangent up to 0.3. Sustained operation is possible up to 0.5, outside of the sweet spot, but generally doesn't provide any significant performance gain in terms of deflection, covered next. Exceeding > 0.5 will lead to thermal runaway.

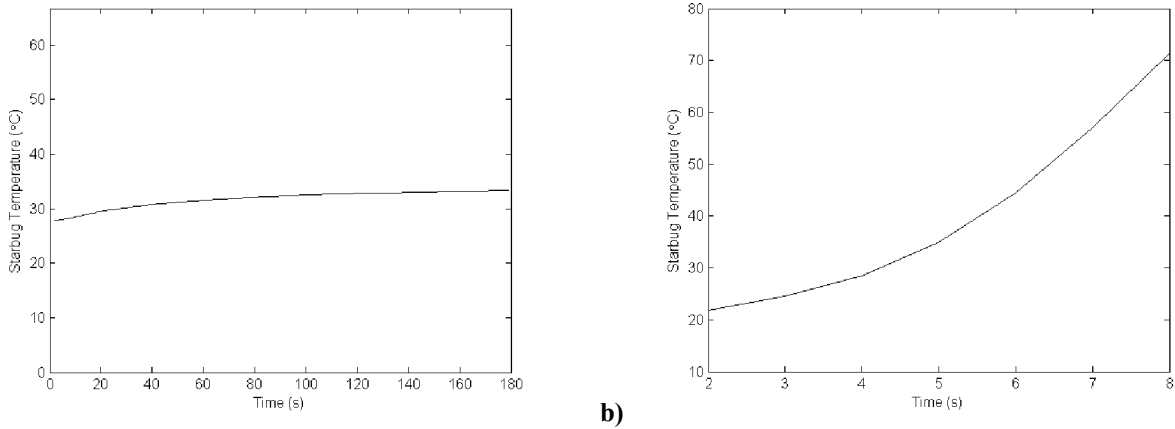

Figure 16: (a) Temperature of Bertha Outer over 180 seconds at 200V/250Hz. The temperature reaches equilibrium in the sweet spot region and can operate continuously without threat of thermal runaway or demagnetization. (b) Temperature of Bertha Outer over 6 seconds at 350V/350Hz, displaying ~Δ45°C in four seconds (4 to 8 seconds).

Table 3: The capacitance is extracted for the data at ($T_{end}$ – 1) seconds, normally 179 seconds to avoid false data. The tabulated voltage/frequency combination in Table 1 for each tube represents the max point of controlled thermal dissipation.

| Tube Name | Temperature (°C) |
|---|---|
| Bertha Outer 1 | 55.8 |
| Bertha Inner 6 | 47.1 |
| Betsy Outer 2 | 48.1 |
| Betsy Outer 4 | 65.6 |
| Belinda Outer 5 | 55.7 |
| Belinda Inner 7 | 43.2 |
| Mohawk Single 3 | 55.0 |

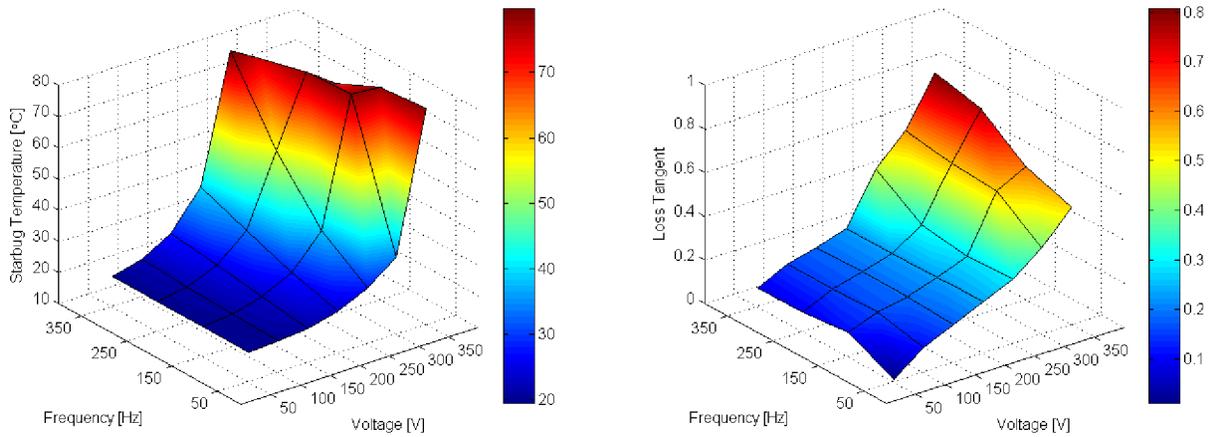

Figure 17: Temperature and loss tangent surface plots created with data from an automated test with Bertha Outer. The sweet spot is in the region up to 250V and 250Hz. The increasing loss tangent with higher voltages/frequencies denotes less effective heat dissipation.

### 5.7 Deflection

The axial displacement of the piezoelectric tube can be modeled as linear for the most part. An increase in voltage causes a proportional increase in axial displacement. Frequency doesn't have an effect on displacement in the sweet spot region. At 250V and 250Hz, a Starbug can buzz happily at a low temperature and minimal power dissipation, with a moderate displacement of 2um, 4um (peak-to-peak). An extra 0.5um, 1um (peak-to-peak) can be squeezed out at 275V/250Hz with the Bertha Outer tube, but the temperature would have to be evaluated thoroughly to ensure the tube can dissipate the additional heat. Going outside the sweet spot, past the high temperature region on the 300V line will cause a degradation of the axial displacement of a piezoelectric tube.

A possible alternative culprit of displacement decay at the high temperature region is the clamp used to secure the tube, which may shift position during higher voltage/frequency tests. There was one instance where a Bertha Outer tube powered at 350V/350Hz vibrated so violently, it shattered in the clamp. Providing some damping material between the tube and the clamp should allow for more tube movement and potentially better displacement measurements at the higher end of the surface plot.

The relative change in length of the piezoelectric tube is roughly -350pm/V ($d_{31}$ models contraction using Eq. 1) on average in the sweet spot for the Bertha Outer tube. The $d_{31}$ constant has a moderate correlation with temperature. Increasing temperature expands the material, which explains why $d_{31}$ increases at higher voltages/frequencies where the piezoelectric material temperature is greater. Using $d_{31}$, an estimate can be made on the relative change in length of the tube without knowing basic tube dimensions. This translates into estimating step size and could be used to calculate lateral displacement, also known as scan range (Δ*x* using Eq. 2). Lateral deflection can be measured using the deflection setup described in this paper, but due to time constraints, was not performed. A metal target and a plastic isolation barrier would have been mounted on the end of one side of the tube. The capacitive probe would then be placed perpendicular to the tube, facing the metal target.

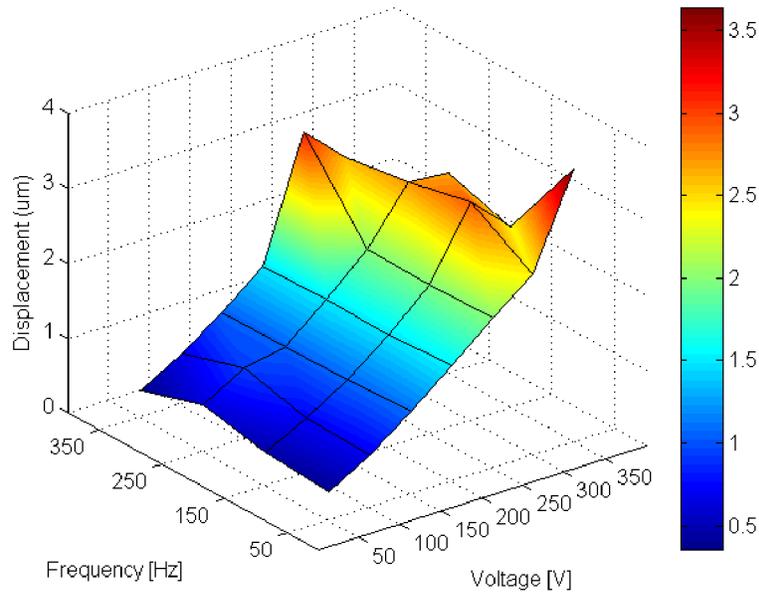

Figure 18: Axial displacement surface graph created with data from an automated test with Bertha Outer. The sweet spot is in the region up to 250V and 250Hz, but can be pushed to 275V (and up to 250Hz) if power dissipation and temperature are not an issue. At high voltages and frequencies, axial displacement degrades.

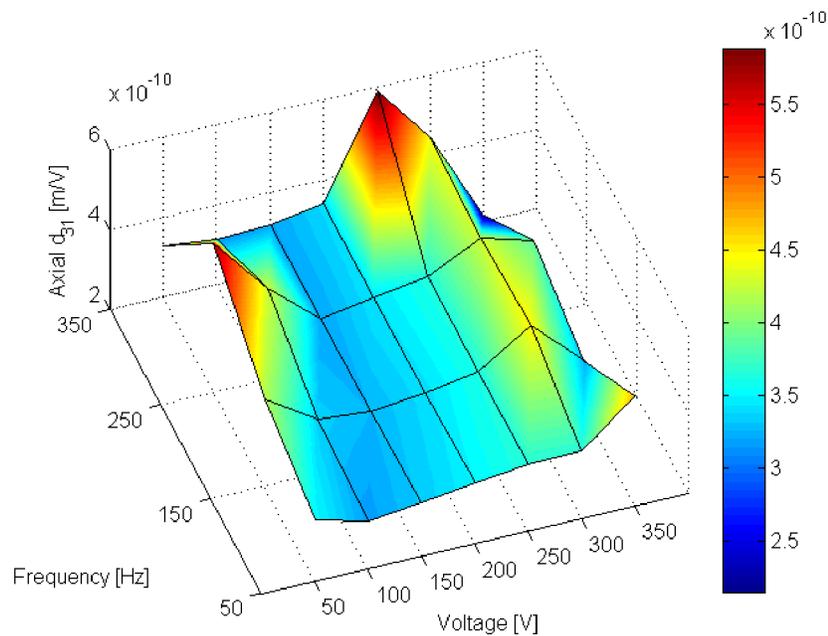

Figure 19: Axial $d_{31}$ surface graph created with data from an automated test with Bertha Outer. The sweet spot is in the region up to 250V and 250Hz, but can be pushed to 275V (and up to 250Hz) if power dissipation and temperature are not an issue. At high voltages and frequencies, the $d_{31}$ constant degrades drastically.

The following equations may be used to calculate $d_{31}$ and lateral deflection, where $d_{31}$ is the piezo constant [m/V]; $L$ is tube length [m]; $\Delta L$ is relative change in tube length during contraction [m]; $\Delta x$ is relative change in lateral displacement (either positive or negative) [m]; $U$ is peak drive voltage [V]; $ID$ is the inner diameter of the tube [m]; and $d$ is the tubes wall thickness [m].

$$\Delta L \approx d_{31} \cdot L \cdot \frac{U}{d} \quad (1)$$

$$\Delta x \approx \frac{\sqrt{2} \cdot d_{31} \cdot L^2 \cdot U}{\pi \cdot ID \cdot d} \quad (2)$$

### 5.8 Active Cooling

The temperature of a piezoelectric tube can affect its overall performance. Increases in temperature can cause large abrupt current draws, leading to considerable power dissipation. If an increase in temperature causes degradation in performance, cooling the piezoelectric tube should lead to improved displacement performance. The following plot compares the peak temperatures of Bertha Outer with and without a cooling fan. The piezoelectric tube was mounted horizontally and suspended in mid-air. The fan was placed directly above the piezoelectric tube, 370mm above. Moderate airflow was directed onto the entire tube.

A quick glance will show that there is significant delta for each frequency in the mid-range (150V – 300V), concluding fan cooling is somewhat effective in the reducing temperature in the sweet spot. The air cooled tube had slightly lower temperatures at higher voltages, but it nonetheless still peaked at the Curie temperature for each point that the non-cooled piezoelectric tube did. Based on this data, the tube cannot be driven at higher voltages with fan cooling in order to boost displacement because fan cooling will not be sufficient to prevent thermal runaway. Other potential cooling methods is pumping chilled air into a chamber containing the tubes, increasing the mass of the Starbugs mounting fitting, or producing a weaker vacuum infused with a minor amount of air in a redesigned subcavity of the Starbug. Pumping chilled air is most probable because it also adds the element of positive pressure, which can help keep dust off the glass field plate used for Starbugs attachment.

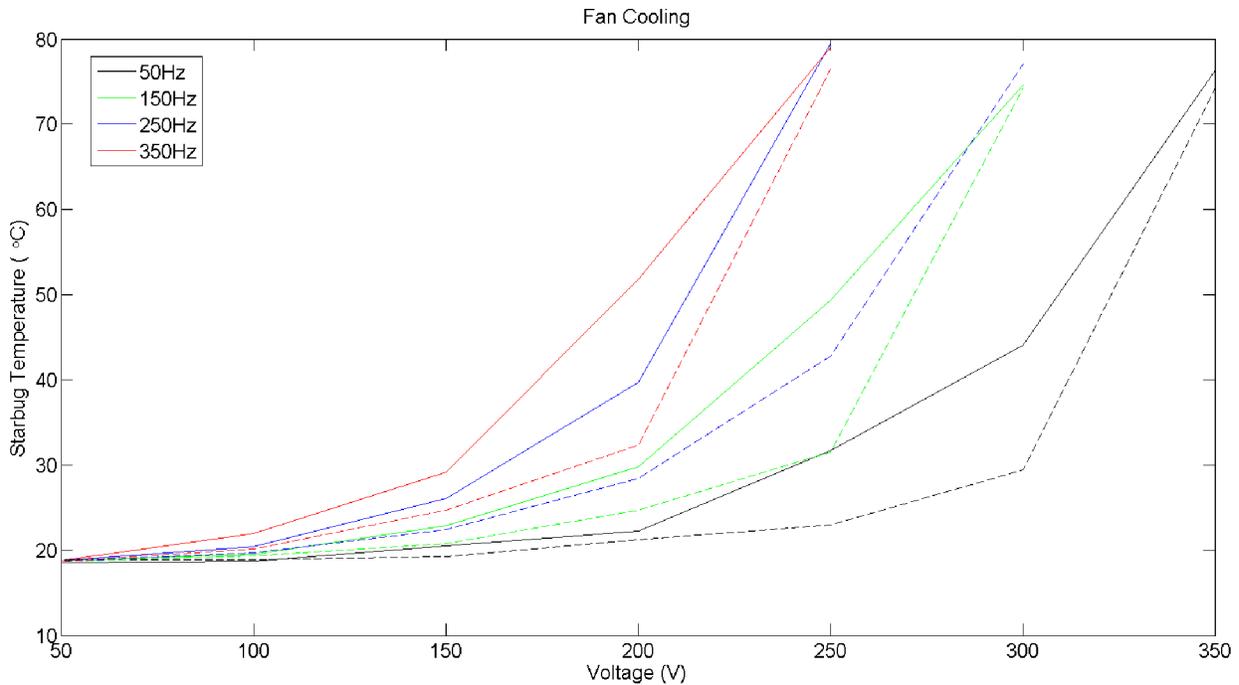

Figure 20: Bertha Outer displaying temperature variation with respect to voltage at four frequencies. Differentiation between fan cooling (segmented lines) and without fan cooling (solid lines) shows large delta in the mid-range voltages. Cooling effects are minimal where either temperature variation is insignificant (≤150V), or exponential (≥ 250V).

## 6. CONCLUSION

An extensive program suite was built with LabVIEW to automate the characterization of Starbugs piezoelectric tubes for power performance, temperature response, and axial/lateral deflection. The acquired data stimulated the critical calculation of the $d_{31}$ piezoelectric constant, allowing direct comparison of piezoelectric tubes as a function of m/V. A region referred to as the 'sweet spot' was found, where key performance characteristics where optimized.

Further testing indicated that imposing aggressive signals on Starbugs piezoelectric tubes, over 250Hz/300V was actually detrimental to performance. The piezoelectric tube power is more sensitive to increases in voltage (with respect to frequency). Leaving the sweet spot led to a 570% power increase (3.5W to 23.5W) with a change of 50V, with negligible improvement to the $d_{31}$ constant for one particular tube. Because the tubes performed better using a lower voltage, the total power consumption is minimized, meaning fewer, less powerful custom-built voltage amplifiers would be required in a full-scale system. Lower power usage provided optimal driving conditions, thus stabilizing the tube temperature to well below the Curie point (demagnetization temperature) of 80°C; this outcome could potentially eliminate the need for active cooling. Axial deflection ($d_{31}$) was found to respond linearly to voltage; there was no significant change/trend with a change in frequency at a fixed voltage. Another key finding is that the lower frequencies used to drive the tubes minimized the mechanical stresses on the brittle ceramic material. The collected data has been exceptional and is proving to be invaluable in the design of future Starbugs for TAIPAN and MANIFEST and other piezoelectric devices.